\documentclass[prb,twocolumn,hyperlinks]{revtex4-1}

\usepackage{graphicx}
\usepackage{mathtools}
\usepackage{color}
\usepackage[colorlinks=true,citecolor=blue]{hyperref}
\usepackage{bbm}

\newcommand{\cumu}[1]{\langle \! \langle #1 \rangle \! \rangle}

\begin{document}

\title{Lee-Yang theory, high cumulants, and large-deviation statistics of the magnetization in the Ising model}

\author{Aydin Deger}
\author{Fredrik Brange}
\author{Christian Flindt}
\affiliation{Department of Applied Physics, Aalto University, 00076 Aalto, Finland}

\date{\today}

\begin{abstract}
We investigate the Ising model in one, two, and three dimensions using a cumulant method that allows us to determine the Lee-Yang zeros from the magnetization fluctuations in small lattices. By doing so with increasing system size, we are able to determine the convergence point of the Lee-Yang zeros in the thermodynamic limit and thereby predict the occurrence of a phase transition. The cumulant method is attractive from an experimental point of view since it uses fluctuations of measurable quantities, such as the  magnetization in a spin lattice, and it can be applied to a variety of equilibrium and non-equilibrium problems. We show that the Lee-Yang zeros encode important information about the rare fluctuations of the magnetization. Specifically, by using a simple ansatz for the free energy, we express the large-deviation function of the magnetization in terms of Lee-Yang zeros. This result may hold for many systems that exhibit a first-order phase transition.
\end{abstract}

\maketitle

\section{Introduction}

In two seminal papers, Lee and Yang investigated phase transitions in many-body systems by considering the zeros of the partition function in the complex plane of the control parameter.\cite{Yang1952a,Lee1952} In particular, they showed how the partition function zeros with increasing system size approach the points on the real axis, where a phase transition occurs. They could thereby explain the non-analytic behavior of the free energy that develops in the thermodynamic limit and signals a phase transition. The Lee-Yang formalism has been applied to a variety of equilibrium problems,\cite{Biskup2000,Bormann2000,Mulken2001,Alves2002,Biskup2004,Bena2005,Lee2013,Lee2013a,GarciaSaez2015,Dijk2015,Krasnytska2015,Krasnytska2016,Gnatenko2017,Deger2018,Giordano2019,Giordano2020} and it has been realized that the framework can also be used to understand non-equilibrium phase transitions, \cite{Arndt2000,Arndt2001,Blythe2002,Blythe2003,Dammer2002,Yin2017} such as dynamical phase transitions in quantum systems after a quench~\cite{Heyl2013,Azimi2016,Heyl2017} and space-time phase transitions in glass formers\cite{Merolle2005,Garrahan2007,Hedges2009,Speck2012} and open quantum systems.\cite{Flindt2013,Hickey2013,Hickey2014}

\begin{figure}[h!]
	\centering
	\includegraphics[width=0.98\columnwidth]{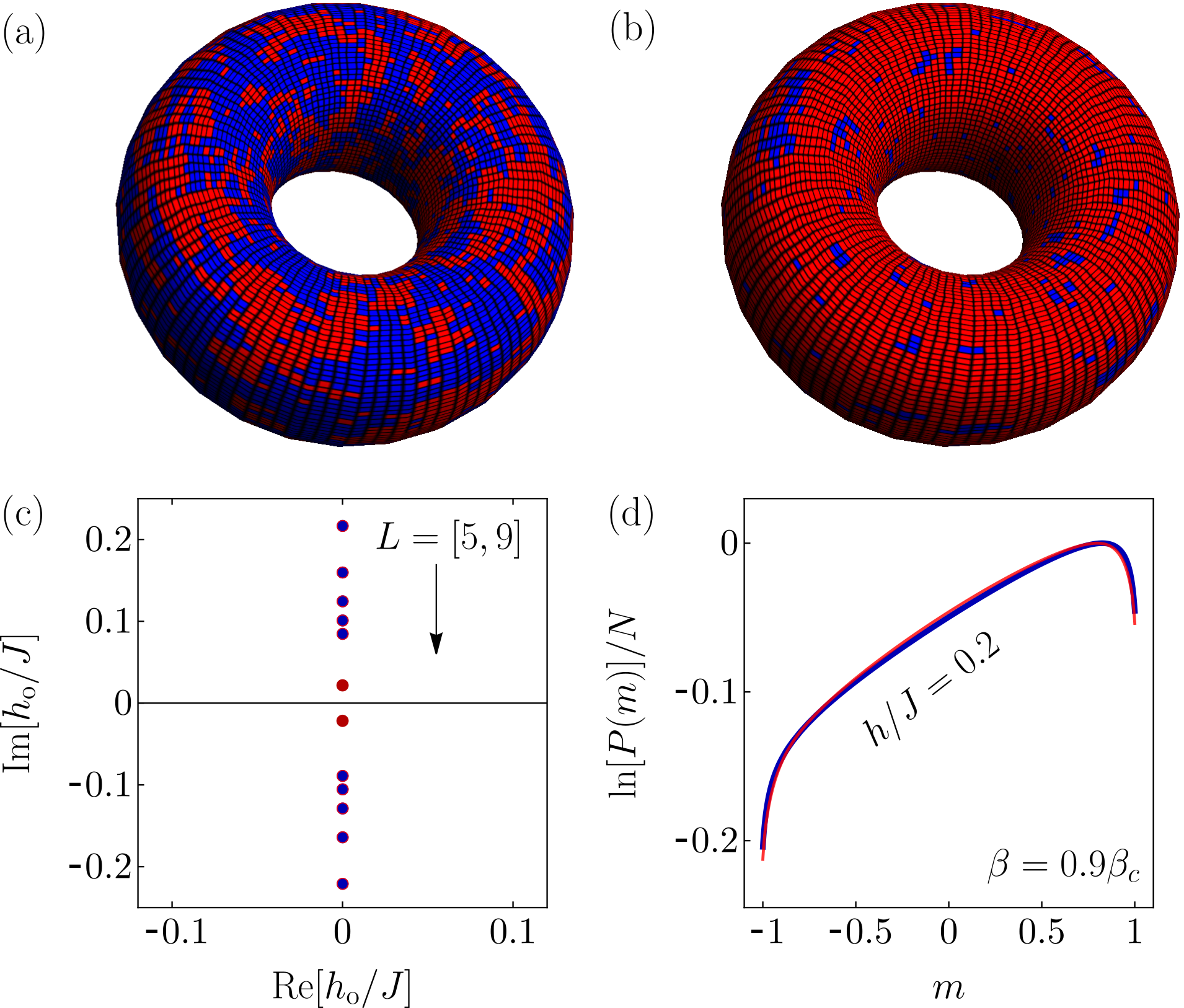}
	\caption{The Ising model, Lee-Yang zeros, and the large-deviation statistics of the magnetization. (a,\,b) The two-dimensional Ising model on a torus with $N=L^d$ spins for $d=2$ and $L=100$. Red (blue) spins point up (down). The temperature is above  the critical temperature, $\beta=0.9 \beta_c$, in the left panel and below it, $\beta=1.1 \beta_c$, in the right panel. No magnetic field is applied, $h=0$. (c) Lee-Yang zeros in the complex magnetic-field plane extracted from the high magnetization cumulants in Ising lattices of linear size $L=5,\ldots,9$, here at the inverse temperature $\beta=0.9 \beta_c$. The red points show the convergence points in the thermodynamic limit. (d) Large-deviation statistics of the magnetization per site, $m=M/N$, for large lattices, $N=L^2\gg 1$. The red curve is numerically exact, while the blue curve is obtained by inserting the convergence points from the left panel into the ansatz~(\ref{eq:LDFansatz}) with $m_0\simeq-1$. }
	\label{fig1}
\end{figure}

In addition to these theoretical developments, partition function zeros have been determined in several recent experiments~\cite{Binek1998,Wei2012,Peng2015,Brandner2017,Flaschner2018} and additional proposals for their detection have been developed.\cite{Wei2017,Gnatenko2018,Kuzmak2019,Krishnan2019} In the approach that we follow here, the partition function zeros are extracted from the fluctuations of the thermodynamic observable that couples to the control parameter, for instance,  magnetization and magnetic field, or energy and inverse temperature.\cite{Flindt2013,Deger2018,Deger2019,Deger2020} The scheme is attractive from an experimental point of view since it makes it possible to explore phase transitions by measuring fluctuations in small systems, even for systems that are away from criticality.\cite{Deger2018,Deger2019,Deger2020} The method was used in the experiment of Ref.~\onlinecite{Brandner2017}, where the dynamical Lee-Yang zeros of an open quantum system were extracted from the statistics of quantum jumps along a stochastic trajectory.\cite{Maisi2014} The partition function zeros are determined from the high cumulants of a fluctuating observable, and the method appears to have a broad scope, since it relies only on a few general properties of partition functions, including an important connection between the zeros of the partition function and its logarithmic derivatives, which deliver the cumulants of interest.\cite{Flindt2013,Deger2018,Deger2019,Deger2020}

The idea is illustrated in Fig.~\ref{fig1} for the two-dimensional Ising model on a torus. From the high cumulants of the magnetization, we extract the Lee-Yang zeros in the complex plane of the magnetic field for small lattices. By doing so with increasing system size, we can determine the convergence points of the Lee-Yang zeros in the thermodynamic limit. An example of this procedure is shown in Fig.~\ref{fig1}(c), where we have extracted the Lee-Yang zeros for the Ising model above the critical temperature. In that case, there is no phase transition, and the Lee-Yang zeros converge to the complex points in red. (Below the critical temperature, they converge to the real axis, as we will see.) Still, the Lee-Yang zeros carry important information about the rare fluctuations of the magnetization, as illustrated in Fig.~\ref{fig1}(d). Here, we compare exact calculations of the large-deviation statistics of the magnetization \cite{Touchette2009} with a simple ansatz given in terms of the extracted Lee-Yang zeros, see Eq.~(\ref{eq:LDFansatz}). The good agreement suggests that a deep connection between Lee-Yang theory and large-deviation statistics may exist. It should be noted that we have considered some of these ideas in recent works on a simple model of a molecular zipper\cite{Deger2018} and the Ising model in a mean-field approximation.\cite{Deger2020} However, to further explore and strengthen these ideas we here consider the Ising model in one, two, and three dimensions as a paradigmatic example of a system that exhibits a phase transition. Thereby we also test and improve our understanding of the cumulant method itself.

The rest of the paper consists of two main parts. Section~\ref{sec:Ising} describes the determination of  Lee-Yang zeros using the cumulant method for the Ising model, while Sec.~\ref{sec:LDSMag} concerns the connection between the Lee-Yang zeros and the large-deviation statistics of the magnetization. Specifically, our work is organized as follows. In Sec.~\ref{subsec:ising}, we introduce the Ising model and the magnetic field zeros of the partition function. In Sec.~\ref{subsec:cumumethod}, we describe the cumulant method, which we use to determine the Lee-Yang zeros from the fluctuations of the magnetization in small lattices. In Sec.~\ref{subsec:LYIsingchain}, we illustrate the method with the Ising model in one dimension, which is a useful example since it is analytically tractable, allowing us to benchmark the extracted Lee-Yang zeros with exact results. In Sec.~\ref{subsec:fini_scaling}, we develop a finite-size scaling analysis of the Lee-Yang zeros, which is needed for the Ising model in higher dimensions. In Sec.~\ref{subsec:Ising lattice}, we apply the cumulant method to the two-dimensional Ising lattice for which we develop a transfer-matrix method for calculating the high cumulants and extracting the Lee-Yang zeros. In Sec.~\ref{subsec:cubicIsing}, we finally determine the Lee-Yang zeros of the Ising model in three dimensions based on Monte-Carlo simulations, which serve to mimic measurements of fluctuations and high cumulants in an experiment. 

In the second part, we consider the large-deviation statistics of the magnetization. In Sec.~\ref{subsec:LDFMag}, we first express the large-deviation statistics in terms of the Lee-Yang zeros based on a simple ansatz for the free energy. In Sec.~\ref{subsec:LDFchain}, we then calculate the large-deviation function for the Ising model in one dimension and show that it indeed can be captured by our ansatz at not too high temperatures. In Sec.~\ref{subsec:squareLDF}, we calculate the large-deviation statistics for the Ising model in two dimensions using a numerically exact transfer-matrix method, and we again find good agreement with the ansatz based on the Lee-Yang zeros. In Sec.~\ref{subsec:cubicLDF}, we use Monte-Carlo simulations for the Ising model in three dimensions to determine the large-deviation statistics of the magnetization. Also in this case, the large-deviation function can be related to the extracted Lee-Yang zeros, indicating that a profound connection between large-deviation statistics and Lee-Yang theory may exist. In Sec.~\ref{sec:conclusions}, we finally summarize our work and provide a perspective on possible developments for the future.

\section{Lee-Yang theory}
\label{sec:Ising}

\subsection{The Ising model \& Lee-Yang zeros}
\label{subsec:ising}

We consider the Ising model of spontaneous magnetization, which describes a lattice of $N=L^d$ spins that take on the values $\sigma_i=\pm1$. Here, we investigate a square lattice of linear size $L$ and dimension~$d$. Figure~\ref{fig1} illustrates the model in two dimensions with periodic boundary conditions corresponding to a torus. An external magnetic field of magnitude $\mathcal{H}$ is applied, and neighboring spins are coupled via a ferromagnetic interaction of strength $\mathcal{J}>0$. The  energy of a spin configuration $\{\sigma_i\}$ reads
\begin{equation}
\mathcal{U}(\{\sigma_i\}) = -\mathcal{J}\sum_{\langle i,j\rangle}\sigma_i\sigma_j-\mathcal{H}\sum_{i}\sigma_i,
\label{eq:Ising_model}
\end{equation}
where the brackets $\langle i,\!j\rangle$ denote summation over nearest-neighbor spins. We also introduce the partition function,
\begin{equation}
Z=\sum_{\{\sigma_i\}} e^{-\beta \mathcal{U}(\{\sigma_i\})},
\label{eq:partition_function}
\end{equation}
where the sum runs over all spin configurations, and $\beta = 1/(k_B T)$ is the inverse temperature. The free energy can be expressed as
\begin{equation}
\mathcal{F}=-\beta^{-1}\ln Z.
\label{eq:free_energy}
\end{equation}
We will also use the following dimensionless quantities, 
\begin{equation}
\begin{split}
J&=\beta\mathcal{J},\\
h&=\beta\mathcal{H},\\
F&=-\beta\mathcal{F},\\
\end{split}
\end{equation}
so that we can express the free energy as
\begin{equation}
F=\ln \Big[\sum_{\{\sigma_i\}} e^{J\sum_{\langle i,j\rangle}\sigma_i\sigma_j+h\sum_{i}\sigma_i}\Big],
\end{equation}
and we write the (dimensionless) free energy per site as
\begin{equation}
f=F/N,
\end{equation}
and the magnetization per site as
\begin{equation}
m=M/N,
\end{equation}
where $M(\{\sigma_i\})=\sum_i \sigma_i$ is the total magnetization for a given spin configuration. The free energy is important to understand the phase behavior of the Ising model. Specifically, in the thermodymic limit, phase transitions are signaled by non-analyticities in the free energy.

To understand this non-analytic behavior, Lee and Yang considered the complex partition function zeros.\cite{Yang1952a,Lee1952} The partition function is an entire function for a finite-size system, as it is a finite sum of exponentials, and it can thus be factorized in terms of its zeros. Considering the partition function as a function of the magnetic field, we can write it as\cite{arfken2012}	
\begin{equation}
\label{eq:parZeros}
Z(h)=Z(0) e^{c h} \prod_k\left(1-h/h_k\right),
\end{equation}
where $h_k$ are the complex magnetic field zeros and $c$ is a constant. The Lee-Yang zeros come in complex conjugate pairs, $h_k$ and $h_k^*$, since the partition function is real for real magnetic fields. The free energy can then be expressed in terms of the Lee-Yang zeros as
\begin{equation}
\label{eq:FparZeros}
F(h)=F(0)+c h +\sum_k\ln(1-h/h_k).
\end{equation}
Lee and Yang showed that the partition function zeros with increasing system size will approach the critical value of the external field for which a phase transition occurs and the free energy becomes non-analytic. These ideas now form the theoretical basis of phase transitions in interacting many-body systems. However, while partition function zeros for a long time were considered a purely theoretical concept, recent works have shown that they can also be determined experimentally.\cite{Binek1998,Wei2012,Peng2015,Brandner2017,Flaschner2018,Wei2017,Gnatenko2018,Kuzmak2019,Krishnan2019}

\subsection{The cumulant method}
\label{subsec:cumumethod}

We now describe the cumulant method that we recently developed to determine the partition function zeros by measuring fluctuations of thermodynamic observables in systems of finite sizes.\cite{Flindt2013,Hickey2013,Hickey2014,Deger2018,Deger2019,Deger2020} Here, we consider the magnetic field zeros, but the method can also be applied to other partition function zeros, for instance, the zeros in the complex plane of the inverse temperature, also known as Fisher zeros. To begin with, we note that the partition sum and the free energy deliver the moments and cumulants of the magnetization upon differentiation with respect to the magnetic field as 
\begin{equation}
\langle M^n\rangle =\frac{\partial_{h}^n Z(h)}{Z(h)},
\label{eq:Mmoment}
\end{equation}
and
\begin{equation}
\langle\!\langle M^n\rangle\!\rangle =\partial_{h}^n F(h).
\label{eq:Mcumulants}
\end{equation}
Importantly, the moments and cumulants of the magnetization can be measured (or obtained from simulations), and the method can thus be experimentally realized to investigate the phase behavior of Ising lattices as well as other interacting many-body systems. Experimentally, cumulants of up to order 15 have been measured for charge transport through a quantum dot.\cite{Flindt2009} To proceed, we differentiate the free energy in Eq.~(\ref{eq:FparZeros}) with respect to the magnetic field and find
\begin{align}  \label{eq:qCumulants}
\langle\!\langle M^{n}\rangle\!\rangle=- \sum_{k} \frac{(n-1)!}{\left(h_{k}-h\right)^{n}}, \quad n>1,
\end{align}
which can be further rewritten in polar coordinates as
\begin{equation}
\langle\!\langle M^{n}\rangle\!\rangle=-(n-1)! \sum_{k} \frac{2\cos\left(n \arg[h_k-h]\right)}{ |h_{k}-h|^{n}}, \quad n>1,
\end{equation}
recalling that the zeros come in complex conjugate pairs. We now see that the sum for large cumulant orders, $n\gg1$, is dominated by the conjugate pair of zeros, $h_{\rm o}$ and $h_{\rm o}^*$,  which are closest to $h$ on the real axis. Thus, for large cumulant orders we can approximate the cumulants as
\begin{equation}
\label{eq:RelationCumZeros}
\langle\!\langle M^{n}\rangle\!\rangle\simeq-(n-1)!\frac{2\cos\left(n \arg[h_{\rm o}-h]\right)}{ |h_{\rm o}-h|^n}, \quad n\gg 1,
\end{equation}
which constitutes an important relation between the leading pair of zeros and the cumulants. In particular, we can invert this expression and determine the leading zeros from the cumulants of the magnetization as \cite{Deger2018, Flindt2013, Deger2019, Deger2020}
\begin{equation}
\label{eq:QMethod}
\begin{bmatrix}
2~\mathrm{Re}\left[h_{\rm o}-h\right]\\
\left|h_{\rm o}-h\right|^2
\end{bmatrix}\simeq\begin{bmatrix}
1& -\frac{\mathsf{\mu}_{n}^{(+)}}{n}\\
1& -\frac{\mathsf{\mu}_{n+1}^{(+)}}{n+1}
\end{bmatrix}^{-1}
\begin{bmatrix}
(n-1) \mathsf{\mu}_{n}^{(-)}\\
n \ \mathsf{\mu}_{n+1}^{(-)}
\end{bmatrix},
\end{equation}
where $\mathsf{\mu}_{n}^{(\pm)} \equiv \langle\!\langle M^{n\pm 1}\rangle\!\rangle / \langle\!\langle M^{n}\rangle\!\rangle$ is the ratio of cumulants of consecutive orders. Inverting the matrix, we find
\begin{widetext}
		\begin{equation}
		\mathrm{Re}\left[h_{\rm o}-h\right]\simeq\frac{n\left(n+1\right) \cumu{M ^n} \cumu{M^{n+1}} -n\left(n-1\right) \cumu{M^{n-1}}   \cumu{M^{n+2}} }{2\left[ (n+1) \cumu{M^{n+1}}^2- \ n \cumu{M^n}  \cumu{M^{n+2}}\right]},  \quad n\gg 1,
		\label{eq:CumulantFormula1}
		\end{equation}
and		
		\begin{equation}
		\left|h_{\rm o}-h\right|^2\simeq\frac{n^2 \left(n+1\right) \cumu{M^n}^2 - n\left(n^2-1\right) \cumu{M ^{n-1}}  \cumu{M^{n+1}}}{(n+1) \cumu{M^{n+1}}^2- \ n \cumu{M^n}  \cumu{M ^{n+2}}},  \quad n\gg 1,
		\label{eq:CumulantFormula2}
		\end{equation}
\end{widetext}
which make it possible to determine the Lee-Yang zeros from the high cumulants of the magnetization.

\begin{figure*}
	\centering
	\includegraphics[width=0.98\textwidth]{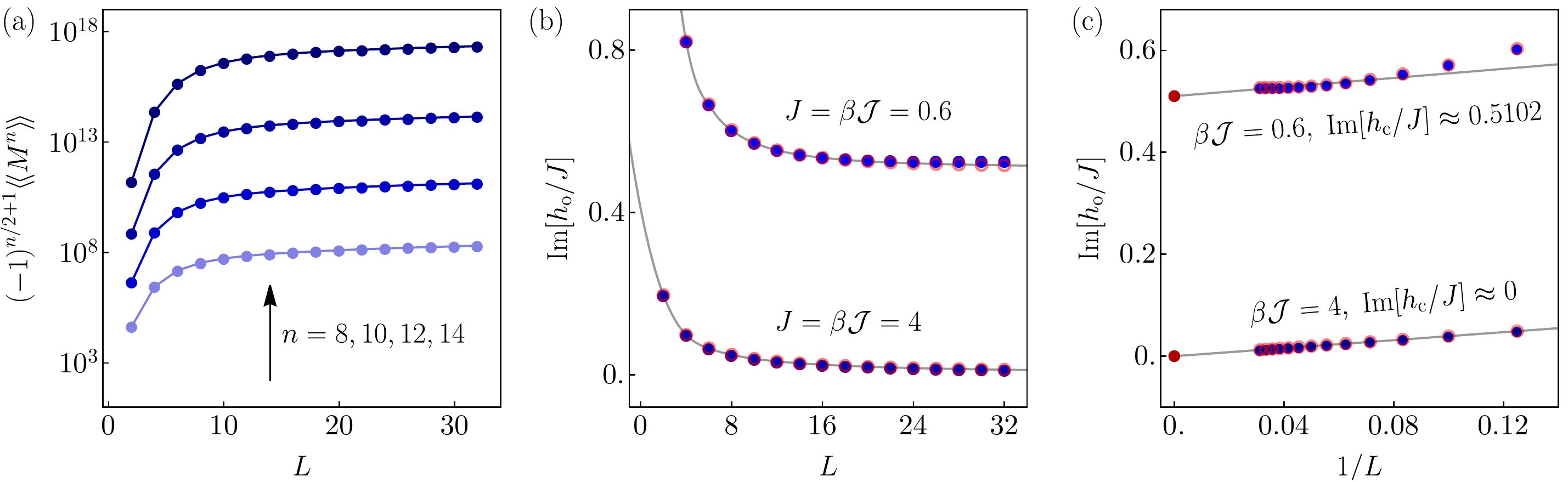}
	\caption{High cumulants of the magnetization and extracted Lee-Yang zeros for the Ising chain. (a) High magnetization cumulants as a function of the system size $L$ at the inverse temperature $\beta \mathcal{J}= 0.6$ and zero magnetic field. The odd cumulants vanish. The lines are guides to the eye. (b) Imaginary part of the Lee-Yang zeros extracted from the cumulants at two different temperatures. The extracted Lee-Yang zeros are shown with filled blue circles, while exact results are indicated with open red circles. The lines are guides to the eye. (c) Determination of the convergence points in the thermodynamic limit.}
	\label{fig2}
\end{figure*}

\subsection{The Ising chain}
\label{subsec:LYIsingchain}

To provide an illustration of the cumulant method, we first consider the one-dimensional Ising lattice with periodic boundary conditions, $\sigma_{L+1}=\sigma_{1}$. In this case, the partition function can be written as
\begin{equation}
Z(h)=\mathrm{Tr}\{\mathbf{T}^L\}=\lambda_+^L+\lambda_-^L
\label{eq:1DZ}
\end{equation}
in terms of the transfer matrix
\begin{equation}
\mathbf{T}=\left(
             \begin{array}{cc}
               e^{J+h} & e^{-J} \\
               e^{- J} & e^{J-h} \\
             \end{array}
           \right)
\end{equation}
and its eigenvalues
\begin{equation}
\lambda_{\pm}=e^{J}\left[\cosh( h)\pm \sqrt{\sinh^2( h)+e^{-4 J}} \right].
\label{eq:1Deig}
\end{equation}
Based on these expressions, it is straightforward to calculate the high cumulants of the magnetization as functions of the system size. We also see that the free energy per site in the thermodynamic limit,
\begin{equation}
f(h)=\ln \max\{\lambda_\pm(h)\},
\label{Free energy per site for 1D Ising}
\end{equation}
is given by the eigenvalue with the largest absolute value. From this expression, it is evident that non-analyticities in the free energy may occur at eigenvalue crossings.

In Fig.~\ref{fig2}a, we show high cumulants of the magnetization as a function of the system size. First, we note that all odd cumulants vanish in the absence of a magnetic field, $h=0$, and that in fact also holds for the Ising model in two and three dimensions. The implementation of the cumulant method then simplifies since Eq.~(\ref{eq:CumulantFormula1}) immediately implies that the real part vanishes, 	$\mathrm{Re}[h_{\rm o}]=0$, and Eq.~(\ref{eq:CumulantFormula2}) reduces to the simple expression 
\begin{equation}
\label{eq:LYZcumulantFormula}
{\rm Im} \left[h_{\rm o}\right]\simeq \pm \sqrt{2n\left(2n+1\right) \left|\frac{\langle \! \langle M^{2n}\rangle \! \rangle}{\langle \! \langle M ^{2(n+1)}\rangle \! \rangle}\right|},\quad n\gg 1,
\end{equation}
involving the ratio of two subsequent even cumulants, which alternate in sign as seen in Fig.~\ref{fig2}a. From this expression, we then determine the Lee-Yang zeros as illustrated in Fig.~\ref{fig2}b for two different temperatures. Finally, to obtain the convergence points in the thermodynamic limit, we extrapolate in Fig.~\ref{fig2}c the position of the Lee-Yang zeros with increasing system size using $1/L$ as a natural small expansion parameter. At finite temperatures, the Lee-Yang zeros remain complex since the Ising chain does not exhibit a thermal phase transition. The zeros only reach the real axis at zero temperature.

To verify the Lee-Yang zeros obtained with the cumulant method, we compare them with exact expressions for the zeros of the partition function. To this end, we solve for the zeros of Eq.~(\ref{eq:1DZ}) and find
\begin{equation}
\ln\lambda_+(h_k)=\ln\lambda_-(h_k)+i \frac{\pi(2k+1)}{L},
\label{eq:eq4zeros}
\end{equation}
where $k$ is an integer. From the explicit expression for the eigenvalues (\ref{eq:1Deig}), we then find the Lee-Yang zeros as
\begin{equation}
h_{k}=\pm i  \arccos\left[\sqrt{1-e^{-4  J}} \cos\left(\frac{\pi(2k+1) }{2 L}\right)\right].
\end{equation}
In Fig.~\ref{fig2}b, we compare the extracted Lee-Yang zeros with the exact results for the leading Lee-Yang zeros ($k=0$) and find very good agreement. We also see that the convergence points in the thermodynamic limit read
\begin{equation}
h_{c}=\pm \ i \arcsin\left(e^{-2 J}\right),
\end{equation}
which again agree well with the results in Fig.~\ref{fig2}c. At finite temperatures, the convergence points remain complex, while at low temperatures, $J=\beta\mathcal{J}\gg1$, we find $h_{c}\simeq \pm \ i e^{-2 J}$, which indeed only vanishes at zero temperature.

\subsection{Finite-size scaling}
\label{subsec:fini_scaling}

To understand the approach of the Lee-Yang zeros to the real axis, we now analyze their finite-size scaling. To this end, we first express the magnetization moments as
\begin{align}\begin{aligned}
\langle M^n\rangle = \int d M \ M^n P(M,L),
\label{eq:defHM}
\end{aligned}\end{align}
where $P(M,L)$ is the probability distribution of the magnetization $M$ for the system of linear size $L$.  To determine the scaling of the moments, we follow the arguments of Binder by invoking a finite-size scaling ansatz for the probability distribution near the critical point,\cite{Kadanoff1966, Domb1983, Binder1981a}
\begin{equation}
P(M, L)=a L^{x}  \tilde{P}\left(b L^{y} M, L / \xi\right).
\label{eq:ansatz}
\end{equation}
Here, the correlation length for large systems is denoted by  $\xi$, the scaling function is denoted by $\tilde{P}$, and $a$, $b$, $x$, and $y$ are constants. The normalization condition, $\int_{-\infty}^{\infty} dM  P(M,L)=1$, translates into the relation~\cite{Binder1981a}
\begin{equation}
L^{x-y} \frac{a}{b} \int_{-\infty}^{\infty} dz  \tilde{P}(z,L/\xi)= 1
\end{equation}
upon the substitution $z= b L^y M$. Since this relation should be valid for any system size, it must hold that $x=y$, and the integral over the scaling function must be constant, such that we can define
\begin{equation}
{c}=\int_{-\infty}^{\infty} dz \tilde{P}(z,0)=\frac{b}{a}, 
\end{equation}
near criticality. We now find the scaling relation
\begin{equation}
\langle M^n \rangle =L^{-n x} f_n(L/\xi),\label{eq:scalingHQ}
\end{equation}
having introduced the function $f_n(L/\xi) =\frac{1}{cb^n} \int dy y^n \tilde{P}(y,L/\xi)$ after the change of variable,~$y\equiv b L^x M$~\cite{Binder1981a}. The following relation, 
\begin{equation}
\langle\!\langle M^n \rangle\!\rangle=\langle M^{n} \rangle-\sum_{m=1}^{n-1}\left(\begin{array}{l}{n-1} \\ {m-1}\end{array}\right) \langle\!\langle M^m \rangle\!\rangle \langle M^{n-m} \rangle,
\label{eq:mom2cumu}
\end{equation}
between cumulants and moments implies that the cumulants must also scale as
\begin{equation}
\label{eq:scalingPhi}
\langle\!\langle M^n \rangle\!\rangle = L^{-n x} ~ g_n(L/\xi),
\end{equation}
where the scaling functions for the cumulants, $g_n$, can be expressed in terms of the scaling functions for the moments, $f_n$. From Eq.~(\ref{eq:qCumulants}), we now see that the zeros must approach the critical value as \cite{Deger2019, Deger2020}
\begin{equation}
	\begin{split}
	\left|h_{\rm o}-h_c\right| &\propto  L^{x},\\
	{\rm Im}[h_{\rm o}] &\propto  L^{x},
	\end{split}
\end{equation}
where $h_c$ (with ${\rm Im}[h_c]=0$) is the critical field for which the system exhibits a phase transition.

\begin{figure*}
	\centering
	\includegraphics[width=0.98\textwidth]{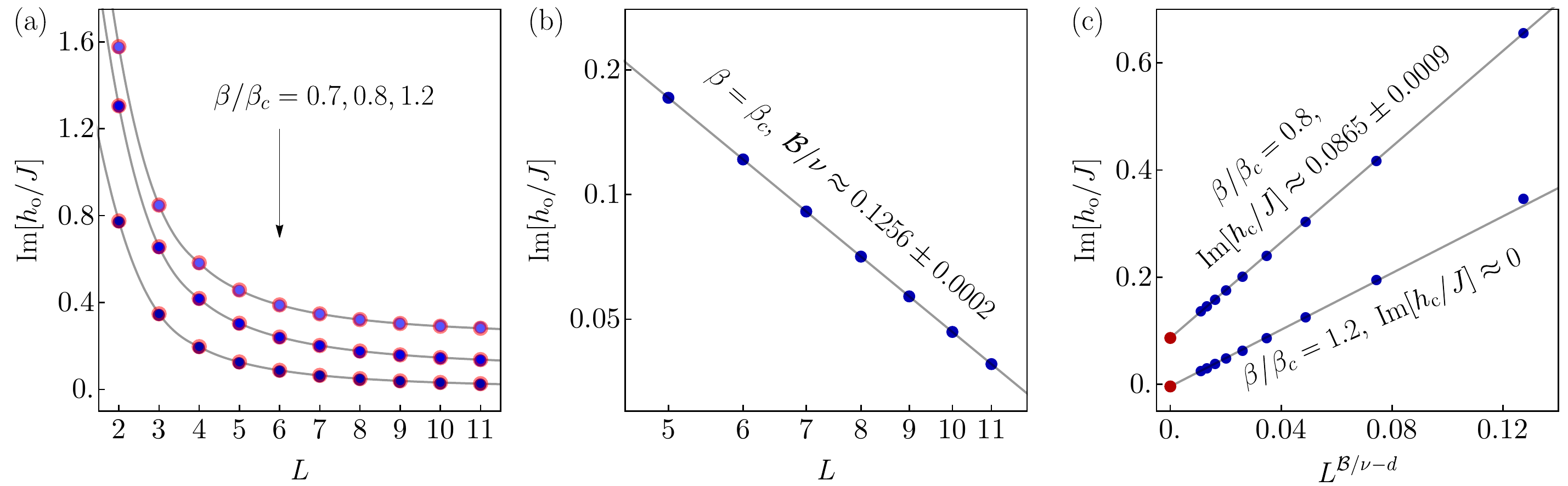}
	\caption{Lee-Yang zeros for the Ising square lattice in two dimensions. (a) Imaginary part of the Lee-Yang zeros, shown with blue points, extracted from the magnetization cumulants with increasing system size for three different temperatures, where $\beta_c = \ln(1+\sqrt{2})/2\mathcal{J}\simeq 0.44/\mathcal{J}$ is the critical inverse temperature. For comparison, we show with red circles numerically exact results for the zeros. (b) Extraction of the ratio of critical exponents, $\mathcal{B}/\nu$, based on Eq.~(\ref{eq:LYscaling}). For the Ising lattice, it is known that $\mathcal{B}=1/8$ and $\nu=1$. (c) Extrapolation of the Lee-Yang zeros in the thermodynamic limit above, $\beta=0.8\beta_c$, and below, $\beta=1.2\beta_c$, the critical temperature. At low temperatures, the Lee-Yang zeros reach the real axis, signaling a phase transition.}
	\label{fig3}
\end{figure*}

In addition, we can relate the critical exponents to the universal critical exponents of the phase transitions. The magnetic susceptibility is defined as 
\begin{equation}
\chi=\partial_h\langle M\rangle/N=\langle\!\langle M^2\rangle\!\rangle/N \propto \xi^{\gamma/\nu},
\end{equation}
where $\gamma$ and $\nu$ are the critical exponents related to the magnetic susceptibility and the correlation length, respectively. We then find
\begin{equation}
\chi= L^{-2x}  g_2(L/\xi)/L^d \propto \xi^{\gamma/\nu},
\label{eq:M2scaling}
\end{equation}
and thereby conclude that $g_2$ must scale as
\begin{equation}
g_2(L/\xi)\propto (L/\xi)^{-\gamma/\nu}.  
\end{equation}
Moreover, the exponents must be related as $2x=-\gamma/\nu-d$ since the right-hand side of Eq.~(\ref{eq:M2scaling}) is independent of~$L$. Drawing on the hyperscaling relation, $\nu d = 2 \mathcal{B}+\gamma$, where $\mathcal{B}$ is the critical exponent associated with the magnetization, we find  $x=\mathcal{B}/\nu-d$. The finite-size scaling for the Lee-Yang zeros then finally becomes
\begin{equation}
	\begin{split}
	\left| h_{\rm o}- h_c\right| &\propto  L^{\mathcal{B}/\nu-d},\\
	{\rm Im}[h_{\rm o}] &\propto  L^{\mathcal{B}/\nu-d}.
	\end{split}
	\label{eq:LYscaling}
\end{equation}
In the next sections, we use these relations to determine the convergence points of the Lee-Yang zeros for the Ising model in two and three dimensions by extrapolating their positions in the thermodynamic limit.

\subsection{The Ising square lattice}
\label{subsec:Ising lattice}

We now consider the Lee-Yang zeros of the two-dimensional Ising model. The model is exactly solvable in the absence of a magnetic field, \cite{Schultz1964} however, here we include the magnetic field to investigate the Lee-Yang zeros. To this end, we write the partition function in terms of a transfer matrix and its eigenvalues as \cite{Huang1987}
\begin{equation}
Z(h)=\mathrm{Tr}\{\mathbf{T}^{L}\} = \sum_{j=1}^{2^L}\lambda^L_j.
\label{sec:Z lattice}
\end{equation}
Here, the $2^L$$\times$\,$2^L$ transfer matrix,
\begin{equation}
\mathbf{T}=[2\sinh(2J)]^{L/2}\mathbf{V}_3\mathbf{V}_2\mathbf{V}_1,
\end{equation}
is given by a product of the three matrices
\begin{equation}
\mathbf{V}_1=\prod_{i=1}^L e^{\Theta \mathbf{X}_i},
\mathbf{V}_2=\prod_{i=1}^L e^{J \mathbf{Z}_i\mathbf{Z}_{i+1}},
\mathbf{V}_3=\prod_{i=1}^L e^{h \mathbf{Z}_i}
\end{equation}
with $\tanh \Theta = e^{-2 J}$, and we have defined
\begin{equation}
\mathbf{X}_i=\mathbbm{1}\otimes\mathbbm{1}\otimes\cdots\otimes\sigma_x\otimes\cdots\otimes\mathbbm{1}\otimes\mathbbm{1},
\end{equation}
and
\begin{equation}
\mathbf{Z}_{i}=\mathbbm{1}\otimes\mathbbm{1}\otimes\cdots\otimes  \sigma_z \otimes\cdots\otimes\mathbbm{1}\otimes\mathbbm{1},
\end{equation}
with the standard Pauli matrices on position $i=1,\ldots,L$,
\begin{equation}
\sigma_x=\left( \begin{array}{cc} 0 & 1 \\ 1 & 0 \end{array} \right),\,\,\,\,
\sigma_z=\left( \begin{array}{cc} 1 & 0 \\ 0 & -1 \end{array} \right).
\end{equation}

In the thermodynamic limit, the free energy per site is determined by the largest eigenvalue of $\mathbf{T}$ as
\begin{align}
f(h)=\frac{1}{L} \ln\max\{\lambda_i(h)\}.
\end{align}
However, in the following we again need to evaluate the cumulants of the magnetization in lattices of finite size. Generally, it is difficult to evaluate high derivatives numerically, and we here pursue a different approach. We express the moments of the magnetization as
\begin{equation}
\langle M^n\rangle = \frac{\mathrm{Tr}\{\partial^n_{h} \mathbf{T}^{L}\}}{Z(h)}
\end{equation}
and then use a recursive expression for the derivatives of powers of the transfer matrix. We first note that
\begin{equation}
\partial_{h} \mathbf{T}^L 	= \mathbf{M}\mathbf{T}^L+ \mathbf{T} \partial_{h} \mathbf{T}^{L-1},
\end{equation} 
having used that $\partial_{h} \mathbf{T} 	= \mathbf{M}\mathbf{T}$, since $\mathbf{V}_3=e^{h \mathbf{M}}$ with 
\begin{equation}
\mathbf{M}=\sum_i \mathbf{Z}_{i}.
\end{equation} 
For the higher derivatives, we then find
\begin{equation}
	\partial^n_{h} \mathbf{T}^{l}= \mathbf{M} \partial^{n-1}_{ h} \mathbf{T}^{l}+ \sum_{m=0}^{n-1}  \binom{n-1}{m}  \mathbf{M}^{n-1-m}\ \mathbf{T} \partial^{m+1}_{ h} \mathbf{T}^{l-1}
\end{equation}
for $l=1,\ldots,L$, having made use of the binomial series,
\begin{equation}
(a+b)^n=\sum_{m=0}^{n} \binom{n}{m} a^{n-m} b^m.
\end{equation}
With these expressions, we can accurately evaluate the high moments of the magnetization and subsequently obtain the cumulants using Eq.~(\ref{eq:mom2cumu}) or, equivalently, using the convenient $n\times n$ determinant formula
	\begin{equation}
	 \frac{\langle\!\langle M^{n} \rangle\!\rangle}{(-1)^{n+1}}=\begin{vmatrix}
	\langle M ^{1}\rangle&1&0&0&\ldots \vspace{2mm}\quad\\
	\langle M ^{2}\rangle&\langle M ^{1}\rangle&1&0&\ldots\vspace{2.5mm}\quad\\
	\langle M ^{3}\rangle&\langle M ^{2}\rangle&\displaystyle\displaystyle\binom{2}{1}\langle M ^{1}\rangle&1&\ldots\vspace{2.5mm}\quad\\
	\langle M ^{4}\rangle&\langle M ^{3}\rangle&\displaystyle\displaystyle\binom{3}{1}\langle M ^{2}\rangle&\displaystyle\binom{3}{2}\langle M ^{1}\rangle&\ldots\vspace{2.5mm}\quad\\
	\ldots&\ldots&\ldots&\ldots&\ddots\quad
	\end{vmatrix}.
	\nonumber
	\end{equation}

In Fig.~\ref{fig3}a, we show the imaginary part of the leading Lee-Yang zeros as a function of the system size $L$, obtained from the high cumulants of the magnetization at three different temperatures, above and below the critical point. In Fig.~\ref{fig3}b, we proceed by using Eq.~(\ref{eq:LYscaling}) to extract the ratio of critical exponents, $\mathcal{B}/\nu$, which comes close to the known value for the two-dimensional Ising model of $1/8$. Finally, in Fig.~\ref{fig3}c, we use the ratio of the critical exponents to extrapolate the position of the Lee-Yang zeros in the thermodynamic limit. Above the critical temperature, the Lee-Yang zeros remain complex, since there is no phase transitions. By contrast, below the critical temperature, the Lee-Yang zeros reach the real axis corresponding to the first-order phase transition that occurs as the magnetic field is tuned across $h=0$. These results show how it is possible to predict the critical behavior of the Ising model by measuring the fluctuations of the magnetization in small lattices.

\begin{figure*}
	\centering
	\includegraphics[width=0.98\textwidth]{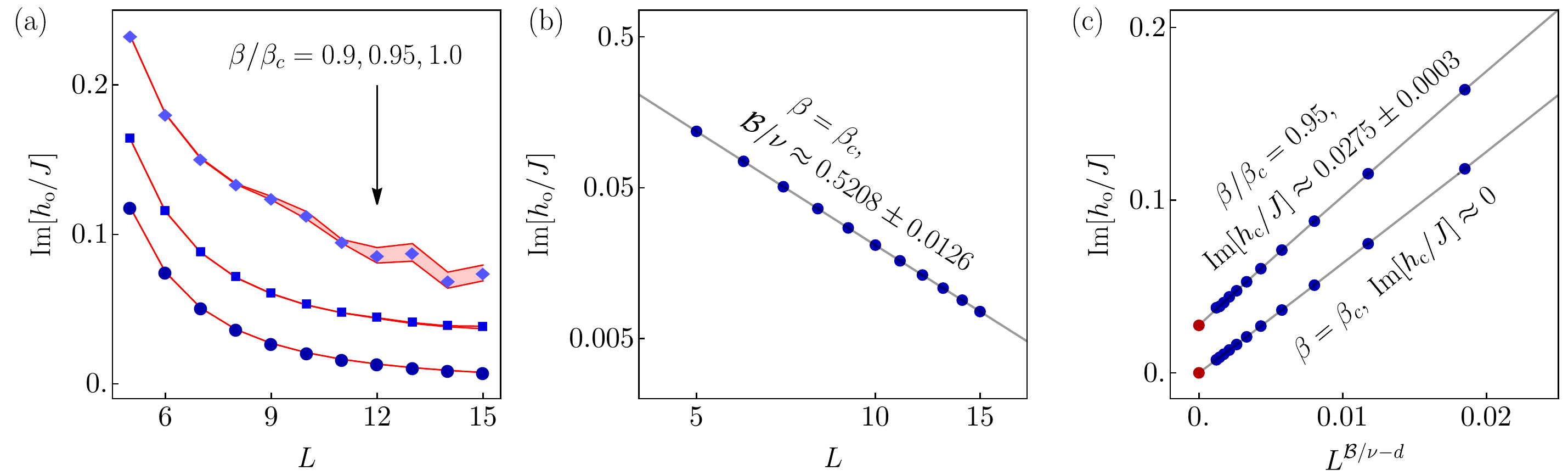}
	\caption{Lee-Yang zeros for the Ising lattice in three dimensions. (a) Imaginary part of the Lee-Yang zeros extracted from the magnetization cumulants at three different temperatures, where $\beta_c\simeq 0.22165/J$ is the critical inverse temperature. The cumulants were obtained by averaging over $5$ sets of Monte-Carlo simulations, each with $5\times10^5$ measurements. The red bands indicate the standard error over the 5 sets. (b) Extraction of the ratio of critical exponents, $\mathcal{B}/\nu$, based on Eq.~(\ref{eq:LYscaling}). The extracted value is close to the best known value of $\mathcal{B}/\nu\simeq 0.5181$, with the uncertainty given by the standard error over the 5 sets. (c) Extrapolation of the Lee-Yang zeros in the thermodynamic limit above and at the critical temperature.}
	\label{fig4}
\end{figure*}

\subsection{The cubic Ising lattice}
\label{subsec:cubicIsing}

The determination of the Lee-Yang zeros in the previous section was based on numerically exact transfer-matrix calculations of the magnetization cumulants. To provide another illustration of the cumulant method, we now consider the cubic Ising lattice in three dimensions for which we determine the magnetization cumulants from Monte-Carlo simulations. In this way, we obtain numerical data for the magnetization cumulants, similarly to what one could measure in an experiment. We can thereby estimate the uncertainty associated with a finite-size sample of data.

\begin{figure*}
	\centering
	\includegraphics[width=0.98\textwidth]{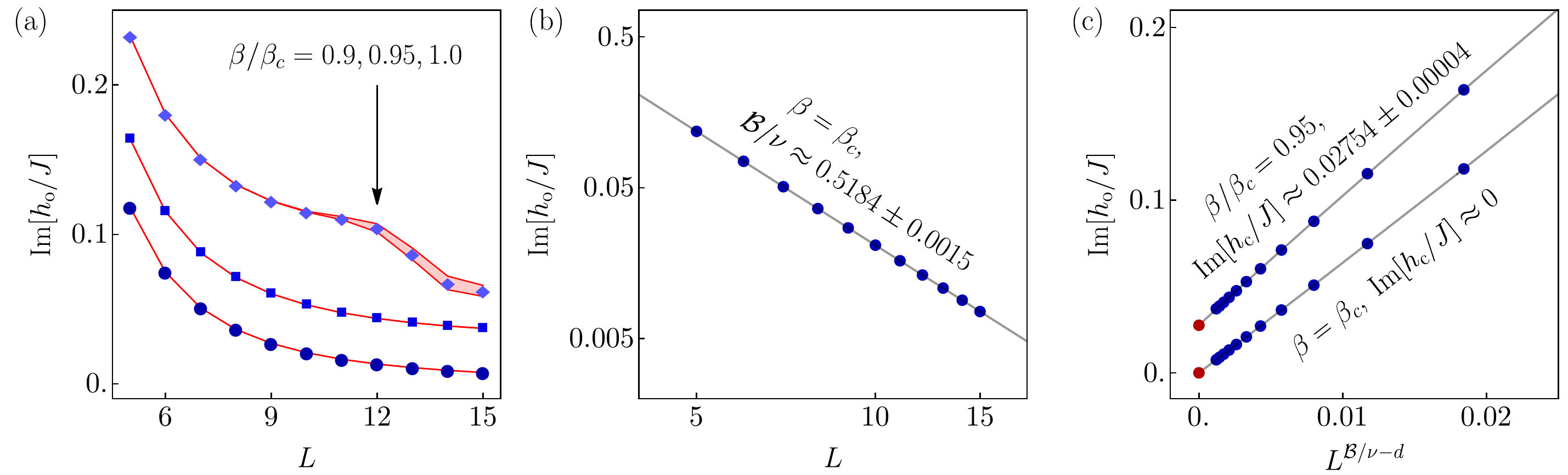}
	\caption{Lee-Yang zeros for the Ising lattice in three dimensions.  (a) Imaginary part of the Lee-Yang zeros extracted from the magnetization cumulants at three different temperatures, where $\beta_c\simeq 0.22165/J$ is the critical inverse temperature. The cumulants were obtained by averaging over $15$ sets of Monte-Carlo simulations, each with $2\times10^7$ measurements. The red bands indicate the standard error over the 15 sets. (b) Extraction of the ratio of critical exponents, $\mathcal{B}/\nu$, based on Eq.~(\ref{eq:LYscaling}). The extracted value is close to the best known value of $\mathcal{B}/\nu\simeq 0.5181$, with the uncertainty given by the standard error over the 15 sets. (c) Extrapolation of the Lee-Yang zeros in the thermodynamic limit above and at the critical temperature.}
	\label{fig5}
\end{figure*}

Figures~\ref{fig4} and~\ref{fig5} show the determination of the Lee-Yang zeros based on the Monte-Carlo simulations. In each figure, we show from left to right the imaginary part of the Lee-Yang zeros as a function of the system size, the determination of the ratio of critical exponents based on Eq.~(\ref{eq:LYscaling}), and finally the extrapolation of the Lee-Yang zeros in the thermodynamic limit. Above the critical temperature, the Lee-Yang zeros remain complex, since there is no phase transition, while they reach the real axis below and at the critical temperature. Importantly, in Fig.~\ref{fig5} we have increased the number of Monte-Carlo simulations compared to Fig.~\ref{fig4}, and we see how the accuracy of the results improve accordingly. These findings illustrate how one could determine the Lee-Yang zeros of a finite-size spin lattice by measuring the fluctuations of the total magnetization.

\newpage

\section{Large-deviation statistics}
\label{sec:LDSMag}

\subsection{Connection to Lee-Yang zeros}
\label{subsec:LDFMag}
In the first part of the paper, we saw how the Lee-Yang zeros can be extracted from the magnetization cumulants in finite lattices. In this second part, we are concerned with the connection between the Lee-Yang zeros and the large-deviation statistics of the magnetization. 
Fundamentally, statistical mechanics and large-deviation theory are intimately linked through concepts such as entropy, rate functions, free energies, and cumulant generating functions.\cite{Touchette2009} Here, we discuss a connection between the Lee-Yang zeros and the large-deviation statistics. To this end, we write the magnetization distribution as
\begin{equation}
P(M)=\sum_{\{\sigma_i\}} \frac{e^{-\beta U(\{\sigma_i\})}}{Z(h)}\int_{-\pi}^\pi \frac{d\chi}{2\pi} e^{i\chi(\sum_{j}\sigma_j-M)},
\label{eq:PM}
\end{equation}
where the Boltzmann factor over the partition function yields the probability for the spin configuration $\{\sigma_i\}$, and we have made use of an integral representation of the Kronecker delta. Using Eq.~(\ref{eq:partition_function}), we then obtain
\begin{equation}
\begin{split}
P(M)&=\int_{-\pi}^\pi \frac{d\chi}{2\pi} \frac{Z(h+i\chi)}{Z(h)}e^{-i\chi M}\\
&=\int_{h-i\pi}^{h+i\pi} \frac{d\kappa}{2\pi i} e^{N[\Theta_m(\kappa)-\Theta_m(h)]},
\end{split}
\label{eq:Mdist}
\end{equation}
having substituted $\kappa = h+i\chi$ and defined the function
\begin{equation}
	\Theta_m(h)=f(h)-mh
\end{equation}
in terms of the free energy and the magnetization per site. For large system sizes, $N\gg1$, the integral in Eq.~(\ref{eq:Mdist}) is amenable to a saddle-point approximation. Specifically, the large-deviation statistics  takes the form
\begin{equation}
\frac{\ln P(m)}{N}\simeq \Theta_m(\kappa_0)-\Theta_m(h),
\label{Saddle-point approximation}
\end{equation}
where $\kappa_0=\kappa_0(m)$ solves the saddle-point equation, $\Theta'_m(\kappa)=0$, which can also be formulated as
\begin{equation}
\langle m\rangle (\kappa) = m.
\label{Saddle point equation}
\end{equation}
Thus, we need to find the value of the auxilary magnetic field $\kappa$ for which the average magnetization would equal~$m$. In many cases, it is difficult to solve Eq.~\eqref{Saddle point equation}, since the free energy and the average magnetization are complicated functions of the magnetic field. 

\begin{figure*}
	\centering
	\includegraphics[width=0.98\textwidth]{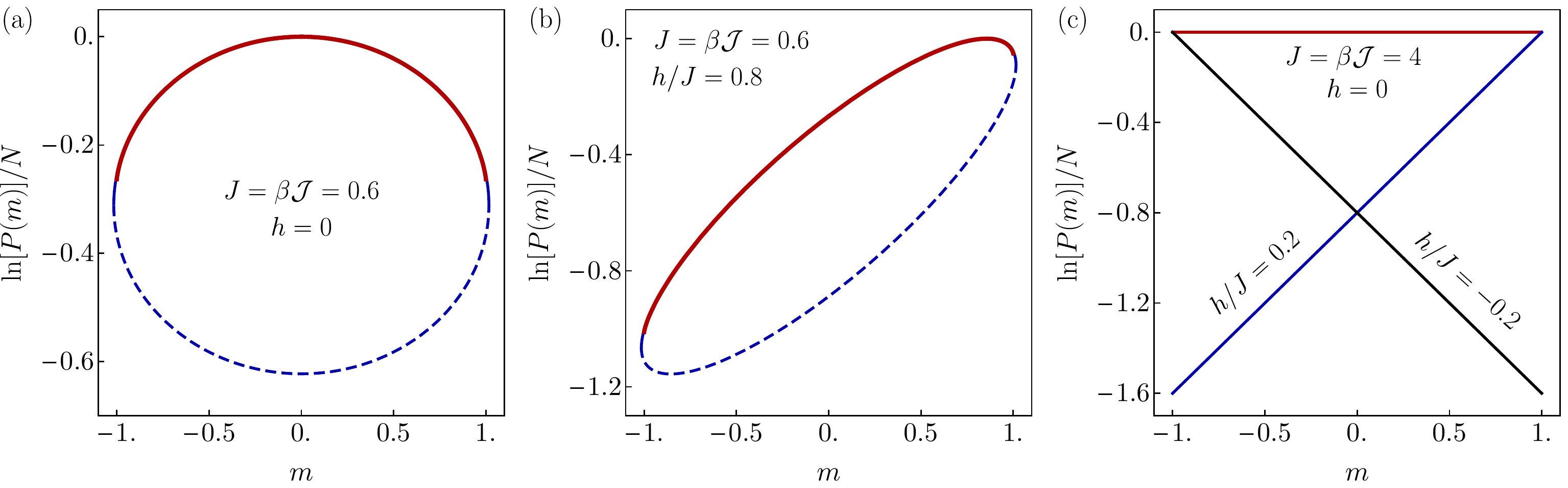}
	\caption{Large-deviation statistics of the magnetization for the Ising chain. (a) Large-deviation statistics at a finite temperature in the absence of a magnetific field. The solid line corresponds to Eq.~(\ref{eq:LDF1D}), while the dashed line is an ellipse whose upper part is given by Eq.~(\ref{eq:LDFansatz}) with the Lee-Yang zeros inserted and $m_0=1$. (b) Large-deviation statistics as in the left panel, but with an applied magnetic field, which tilts the ellipse. (c) At very low temperatures, the ellipse collapses to a nearly straight line.}
	\label{fig6}
\end{figure*}

To find an approximate solution of the saddle-point equation, we now make a crude ansatz for the free energy. Specifically, we assume that the Lee-Yang zeros converge to square-root branch points of the free energy, which is typical for eigenvalue crossings. Thus, close to the convergence points, $h_c$ and $h_c^*$, we make the ansatz~\cite{Deger2018,Deger2020}
\begin{equation}
\label{eq:ansatz0}
f(h)\simeq m_0\sqrt{\left(h_{c}-h\right)\left(h_{c}^{*}-h\right)}
\end{equation}
for the free energy, where $m_0$ is an unknown parameter.
For the average magnetization, we now find
\begin{equation}
\langle m \rangle(h) \simeq \frac{m_0\big[h-{\rm R e}\left(h_{c}\right)\big]}{\sqrt{\left|h_{c}\right|^{2}+h\left[h-2 {\rm Re} \left(h_{c}\right)\right]}},
\end{equation}
and we can then solve Eq.~(\ref{Saddle point equation}), which yields
\begin{equation}
\kappa_{0} \simeq {\rm{Re}}({h_c}) + |{\rm{Im}}\left(h_{\mathrm{c}}\right)|\frac{m}{\sqrt{m_0^2-m^2}},
\end{equation}
having chosen the solution for which the average magnetization increases as the magnetic field is increased, taking the imaginary part of $h_c$ to be positive. (We throw away a solution for which the magnetization decreases.) Inserting this solution into the ansatz for the free energy, we find a  simple expression for the large-deviation statistics,
\begin{equation}
\frac{\ln P(m)}{N}\simeq  m[h-\mathrm{Re}(h_c)]-f(h)+|{\rm{Im}}\left(h_{\mathrm{c}}\right)|\frac{m_0|m_0|-m^2}{\sqrt{m_0^2-m^2}}.
\label{eq:LDFansatz}
\end{equation}

From the approximation above, we expect that the large-deviation statistics will simply be given by a straight line with slope $h-\mathrm{Re}(h_c)$, if the Lee-Yang zeros reach the real axis and the convergence point thus is real. The third term, $f(h)$, does not involve the magntization and is just a constant vertical shift of the large-deviation function. The last term becomes relevant, if the Lee-Yang zeros do not reach the real axis, and the convergence points remain complex. Interestingly, this term is independent of the magnetic field. Thus, if the large-deviation function is known at zero magnetic field, one can predict how it will evolve as a magnetic field is applied. Furthermore, if $m_0$ is positive, the last term simplifies and then describes the upper part of an ellipse.

The expression for the large-deviation function provides a link between Lee-Yang theory and large-deviation statistics. However, it relies on the ansatz~(\ref{eq:ansatz0}) for the free energy, which is not obvious, although it may capture many essential features of first-order phase transitions. To improve our understanding of the ansatz, we now compare it with the large-deviation statistics for the Ising model in one, two, and three dimensions.

\subsection{The Ising chain}
\label{subsec:LDFchain}

For the Ising chain, we can solve Eq.~\eqref{Saddle point equation} exactly using the explicit expression for the free energy in Eq.~\eqref{Free energy per site for 1D Ising}. The saddle-point equation then reads
\begin{equation}
\langle m\rangle (\kappa)=\frac{ \sinh (\kappa )}{\sqrt{\sinh ^2(  \kappa )+e^{-4  J}}} = m
\end{equation} 
with the solution
\begin{align}
\kappa_{0}= \sinh ^{-1}\left(\frac{m}{\sqrt{1-m^2}}e^{-2J}\right).
\label{eq:kappa0}
\end{align}
Inserting this solution  into Eq.~\eqref{Saddle-point approximation}, we find the large-deviation function for the Ising chain,
\begin{equation}
\begin{split}
\frac{\ln P(m)}{N}=&mh-\ln \left[\cosh (h)+\sqrt{\sinh ^2(h)+e^{-4 J}}\right]\\
&+\ln \left[\frac{e^{-2  J}}{\sqrt{1-m^2}}+\sqrt{1+\frac{m^2}{1-m^2}e^{-4J} }\right]\\
&-m\sinh ^{-1}\left(\frac{m }{\sqrt{1-m^2}}e^{-2  J}\right).
\end{split}
\label{eq:LDF1D}
\end{equation}
At first sight, the result does not resemble Eq.~(\ref{eq:LDFansatz}). However, upon closer inspection, we see that at low temperatures, $e^{-2J}= e^{-2\beta\mathcal{J}}\ll 1$, we can expand it as
\begin{equation}
\frac{\ln P(m)}{N}\simeq mh-f(h)+e^{-2J}\sqrt{1-m^2},
\end{equation}
which exactly corresponds to Eq.~(\ref{eq:LDFansatz}) with $h_c=\pm i \arcsin (e^{-2J})\simeq \pm i e^{-2J}$ and $m_0=1$. At even lower temperatures,  $J=\beta\mathcal{J}\gg1$, it simplifies further to the straight line,
\begin{equation}
\frac{\ln P(m)}{N}\simeq mh-|h|.
\end{equation}

Figure \ref{fig6} shows the large-deviation statistics of the Ising chain for different temperatures and magnetic fields. We show the exact result in Eq.~(\ref{eq:LDF1D}) together with the approximation in Eq.~(\ref{eq:LDFansatz}), where we have inserted the convergence points of the Lee-Yang zeros that we extracted from the magnetization cumulants. The figure illustrates how the ansatz for the free energy in Eq.~(\ref{eq:ansatz0}) leads to an accurate description of the large-deviation function, and it provides an important link between Lee-Yang theory and large-deviation statistics. Next, we explore this link in further detail for the Ising model in higher dimensions.

\begin{figure*}
	\centering
	\includegraphics[width=0.98\textwidth]{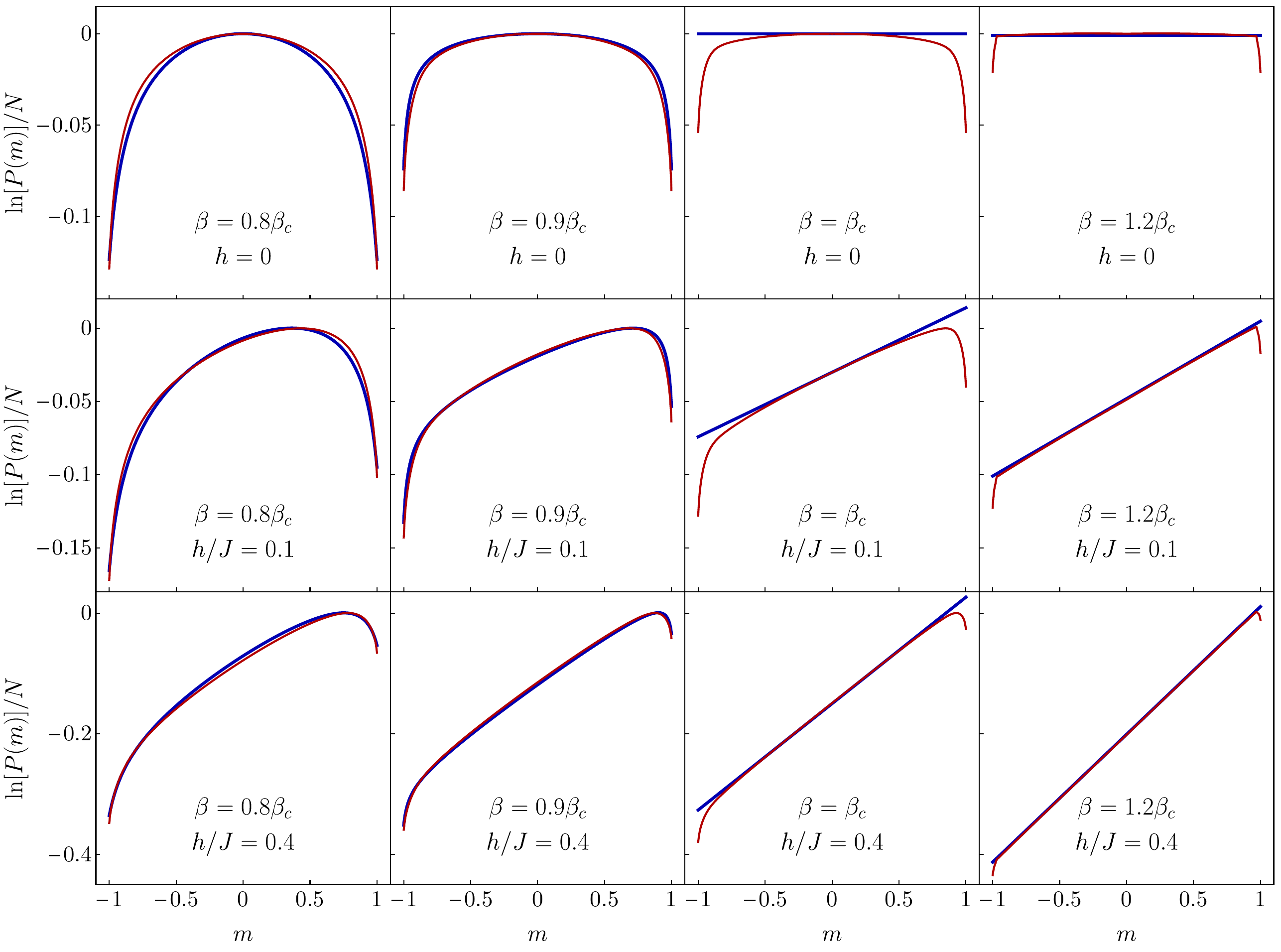}
	\caption{Large-deviation statistics for the two-dimensional Ising lattice. Numerically exact results are shown in red, while the blue lines are the approximation~(\ref{eq:LDFansatz}) with the convergence points of the Lee-Yang zeros inserted, having used $m_0=-\left(1+{\rm Im}[h_c/J]\right)$, which provides a good approximation. For $\beta=0.8\beta_c$, the convergence point is ${\rm Im}[h_c/J]=0.0865(9)$ and  $m_0\approx-1.0865$, while for $\beta=0.9\beta_c$, we have ${\rm Im}[h_c/J]=0.0209(5)$ and $m_0\approx-1.0209$. In the last two columns, we have ${\rm Im}[h_c/J]=0$.
	}
	\label{fig7}
\end{figure*}

\subsection{The Ising square lattice}
\label{subsec:squareLDF}

For the Ising square lattice, we calculate the large-deviation statistics using a numerically exact approach. For large lattices, we can write the average magnetization per site as the logarithmic derivative of the largest eigenvalue of the transfer matrix,
\begin{equation}
\langle m \rangle = \partial_{h} f(h)=  \frac{\partial_{h} \lambda_{\rm max}}{L\lambda_{\rm max}}
\label{eq:magPSinf},
\end{equation}
where the eigenvalue problem reads
\begin{align}
\label{eq:Teigen}
\mathbf{T} |\lambda_{\rm max}\rangle=\lambda_{\rm max} |\lambda_{\rm max}\rangle,
\end{align}
and the left and right eigenvectors are normalized as $\langle\lambda_{\rm max}|\lambda_{\rm max}\rangle=1$.
Now, using the Hellmann-Feynman theorem, we can express the magnetization as
\begin{equation}
\begin{split}
\langle m \rangle &=  \frac{\langle\lambda_{\rm max}|(\partial_{h}\mathbf{T})|\lambda_{\rm max}\rangle}{L\lambda_{\rm max}}\\
&=\frac{\langle\lambda_{\rm max}|\mathbf{M} \mathbf{T}|\lambda_{\rm max}\rangle}{L\lambda_{\rm max}}\\
&= \langle\lambda_{\rm max}|\mathbf{m}|\lambda_{\rm max}\rangle, \quad \mathbf{m}=\mathbf{M}/L,
\end{split}
\end{equation}
which resembles the expectation value of an observable in quantum mechanics. Based on this result, we can numerically calculate the magnetization as a function of the magnetic field and then solve the saddle-point equation using a standard numerical root-finding method.

In Fig.~\ref{fig7}, we show the large-deviation statistics of the magnetization for the Ising square lattice in two dimensions. Results are displayed for several different temperatures, above, below, and at the critical temperature, both with and without an applied magnetic field. Just as for the Ising chain, the exact results for the large-deviation statistics are very well captured by the approximation~(\ref{eq:LDFansatz}), where we have inserted the convergence points of the Lee-Yang zeros. Below and at the critical temperature, the large-deviation statistics is given by a straight line, whose tilt is determined by the applied magnetic field. In particular, as the magnetic field is tuned across zero, the average magnetization (given by the highest point on the curves) displays an abrupt jump, corresponding to a first-order phase transition. By contrast, above the critical temperature, the Lee-Yang zeros remain complex, and the imaginary part of the zeros gives rise to the finite curvature of the large-deviation statistics. Again, an applied magnetic field tilts the distributions, however, the change of the average magnetization is smooth, since there is no phase transition in this case.

\begin{figure*}
	\centering
	\includegraphics[width=0.98\textwidth]{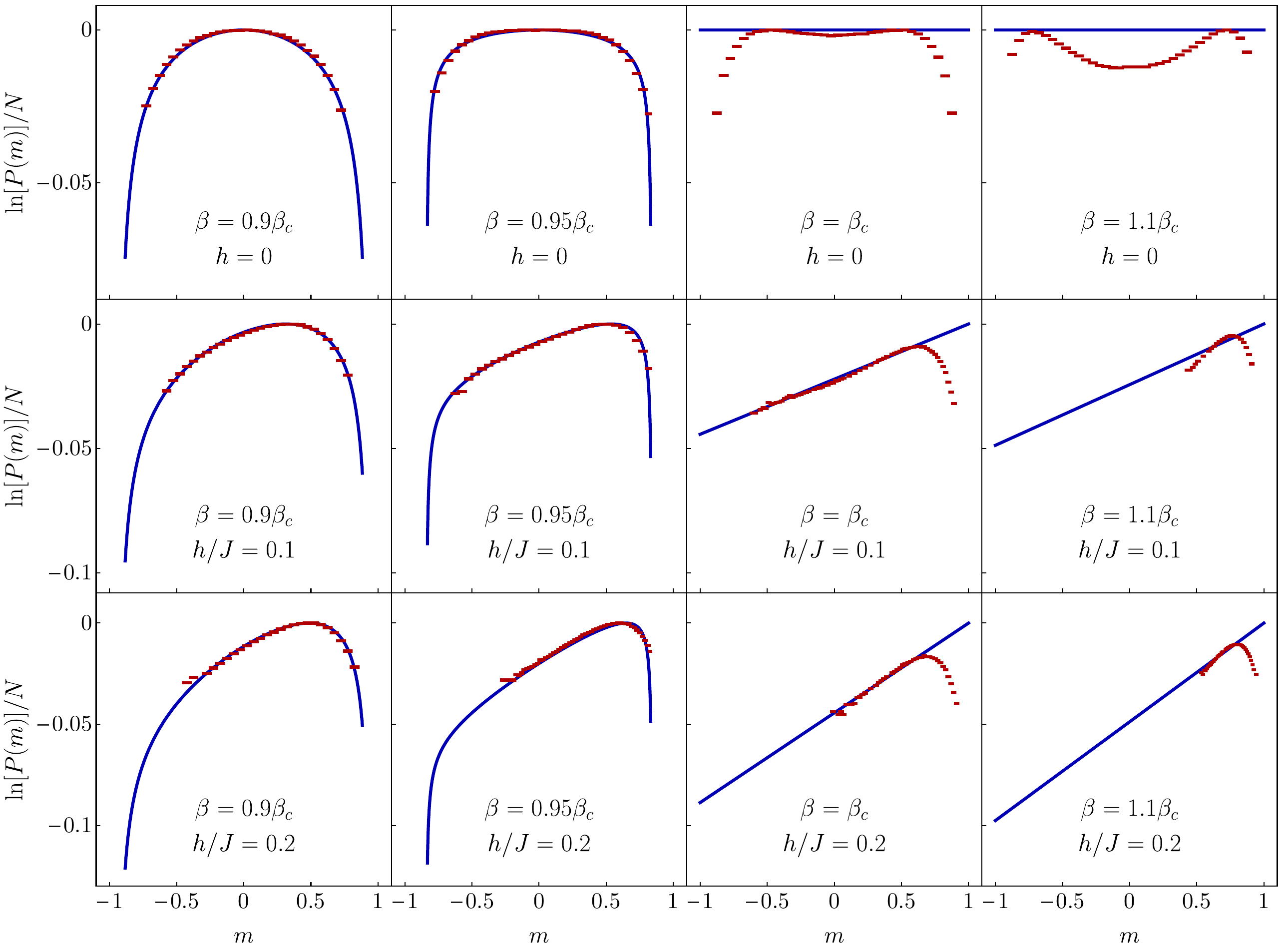}
	\caption{Large-deviation statistics for the three-dimensional Ising lattice. Results based on Monte-Carlo simulations are shown in red, while the blue lines are the approximation~(\ref{eq:LDFansatz}) with the convergence points of the Lee-Yang zeros inserted. The width of the red markers indicate the bin size. In the first three columns, we have used the linear size $L=8$, and in the fourth one $L=10$. For the first two columns, the Lee-Yang zeros converge to the complex points ${\rm Im}[h_c/J]=0.0876(4)$ and  ${\rm Im}[h_c/J]=0.0275(4)$, respectively, and we have used $m_0=-0.94$ and $m_0=-0.84$ for the fitting. In the last two columns, the Lee-Yang zero converge to $h_c=0$. For the Monte-Carlo simulations, we have used $2\times 10^7$ measurements for each panel.}
	\label{fig8}
\end{figure*}

\subsection{The cubic Ising lattice}
\label{subsec:cubicLDF}

Finally, we return to the cubic Ising lattice in three dimensions and evaluate the large-deviation statistics using Monte-Carlo simulations. Here, the key challenge is to accurately sample the rare events in the tails of the distributions for sufficiently large lattices. The results of the Monte-Carlo simulations are shown in Fig.~\ref{fig8} for different temperatures, above, below, and at the critical point. We also show results with an applied magnetic field, which tilts the large-deviation statistics of the magnetization. Above the critical temperature, the distribution already takes on the large-deviation form with $N=8^3=512$ spins, and the results are captured by the ansatz~(\ref{eq:LDFansatz}) with the complex convergence points of the Lee-Yang zeros inserted. From the Monte-Carlo simulations, we can accurately sample the bulk part of the distributions, however, we cannot access the very rare events in the tails due to the limited number of measurements. As we move the critical point in the third column, we see that the distribution of the magnetization has not completely reached the large-deviation form, which should be concave.\cite{Touchette2009} Thus, the slightly bimodal distribution reveals that the system is not quite large enough to take on the large-deviation form. Still, the part of the distribution that can be accessed with the Monte-Carlo simulations is well-captured by the ansatz~(\ref{eq:LDFansatz}). Finally, in the last column, where the temperature is below the critical point, we have increased the system size to $N=10^3=1000$ spins. In this case, however, this system is still not large enough to take on the large-deviation form and clear finite-size effects are visible. The increased system size also makes it further difficult to access the rare tails of the distribution, which are exponentially suppressed with the system size, and therefore exponentially harder to realize in the Monte-Carlo simulations. On the other hand, using the ansatz~(\ref{eq:LDFansatz}) together with the convergence points of the Lee-Yang zeros, we can predict the exponentially small probabilities to observe a rare fluctuation of the magnetization, as indicated by the blue line. It is worth to mention that more advanced Monte-Carlo methods exist,\cite{Newman2001,Landau2014,Tsypin2000} such as flat-histogram methods, umbrella sampling, and multicanonical methods, however, implementing them is beyond the scope of this work.

\section{Conclusions}
\label{sec:conclusions}

In summary, we have used a recently developed cumulant method to determine the Lee-Yang zeros of the Ising model in one, two and three dimensions from the high cumulants of the magnetization in lattices of finite size. The method is based on the fluctuations of the magnetization, which in principle can be measured, and our approach is therefore attractive from an experimental point of view. Having determined the convergence points of the Lee-Yang zeros in the thermodynamic limit, we have shown how they encode important information about the large-deviation statistics of the magnetization. In particular, using a simple ansatz for the free energy, we have expressed the large-deviation function in terms of the Lee-Yang zeros and found good agreement with calculations for the Ising model in all three dimensions. This result may hold for many systems that exhibit a first-order phase transition.

Our work opens several perspectives for future research. While the determination of Lee-Yang zeros here was based on calculations of the magnetization cumulants, it would be interesting to implement our method in an experiment by measuring the magnetization fluctuations in an Ising lattice of finite size. In addition, the method is not restricted to equilibrium problems only, but may equally well be applied to dynamical phase transitions in quantum many-body systems after a quench\cite{Heyl2013,Azimi2016,Heyl2017} or quantum phase transitions in the groundstate of an interacting quantum spin chain.\cite{Lamacraft2008,Xu2019}

\begin{acknowledgments}
We acknowledge the computational resources provided by the Aalto Science-IT project. A.D.~acknowledges support from the Vilho, Yrjö and Kalle Väisälä Foundation of the Finnish Academy of Science and Letters through the grant for doctoral studies. F.B. acknowledges support from the European Union’s Horizon 2020 research and innovation programme under the Marie Skłodowska--Curie grant agreement No.~892956. The work was supported by the Academy of Finland (Projects No. 308515, 312057, 312299, and 331737).
\end{acknowledgments}

\nocite{apsrev41Control}
\bibliographystyle{apsrev4-1}

\end{document}